\documentclass[lettersize,journal]{IEEEtran}
\usepackage{amsmath,amsfonts}
\usepackage{algorithmic}
\usepackage{algorithm}
\usepackage{array}
\usepackage[caption=false,font=normalsize,labelfont=sf,textfont=sf]{subfig}
\usepackage{textcomp}
\usepackage{stfloats}
\usepackage{url}
\usepackage{verbatim}
\usepackage{graphicx}
\usepackage{cite}
\usepackage{caption}
\usepackage{cite}

\begin{document}

\title{Caching Techniques for Reducing the Communication Cost of Federated Learning in IoT Environments}

\author{
\begin{center}
\begin{minipage}{0.45\textwidth}
    \centering
    \textbf{Ahmad Alhonainy} \\
    \textit{Department of Electrical Engineering} \\
    \textit{\& Computer Science} \\
    \textit{The University of Missouri} \\
    Columbia, USA \\
    aga5h3@umsystem.edu
\end{minipage}
\hfill
\begin{minipage}{0.45\textwidth}
    \centering
    \textbf{Praveen Rao} \\
    \textit{Department of Electrical Engineering} \\
    \textit{\& Computer Science} \\
    \textit{The University of Missouri} \\
    Columbia, USA \\
    praveen.rao@missouri.edu
\end{minipage}
\end{center}
}

\maketitle

\begin{abstract}
Federated Learning (FL) allows multiple distributed devices to jointly train a shared model without centralizing data, but communication cost remains a major bottleneck, especially in resource-constrained environments. This paper introduces caching strategies—FIFO, LRU, and Priority-Based—to reduce unnecessary model update transmissions. By selectively forwarding significant updates, our approach lowers bandwidth usage while maintaining model accuracy. Experiments on CIFAR-10 and medical datasets show reduced communication with minimal accuracy loss. Results confirm that intelligent caching improves scalability, memory efficiency, and supports reliable FL in edge-IoT networks, making it practical for deployment in smart cities, healthcare, and other latency-sensitive applications. 
\end{abstract}


\section{Introduction}
\IEEEPARstart{F}{ederated} Learning (FL)~\cite{mcmahan2017communication} is a decentralized machine learning paradigm that enables edge devices to collaboratively train a global model without sharing their raw data. Instead, each client computes model updates locally and sends them to a central server for aggregation and global model enhancement. This approach enhances privacy and reduces data transmission but introduces substantial communication overhead—especially in bandwidth-constrained Internet of Things (IoT) environments.

\noindent Formally, FL aims to minimize a global objective:

\[
\min_{\theta \in \mathbb{R}^d} F(\theta) = \sum_{i=1}^{N} \frac{n_i}{n} F_i(\theta),
\] where \( F_i(\theta) \) is the local loss function for client \( i \), based on its private dataset of size \( n_i \), and \( n = \sum_{i=1}^{N} n_i \). 

This paper proposes a caching-based communication optimization framework for FL. We selectively filter and store client updates at the server using lightweight strategies—First-In-First-Out (FIFO)~\cite{pratheeksha2021mlcache}, Least Recently Used (LRU)~\cite{pratheeksha2021mlcache}, and Priority-Based Replacement (PBR)—to reduce redundancy and lower communication cost without degrading model accuracy.

\noindent Key contributions include:

\begin{itemize}
    \item A dynamic threshold mechanism to filter insignificant updates;
    \item Three cache replacement strategies for memory-aware update selection;
    \item Experimental validation on CIFAR-10~\cite{krizhevsky2009learning} and medical datasets showing reduced communication (up to 20\%) and maintained or improved accuracy.
\end{itemize}




\section{Related Work}

FL enables decentralized training while preserving user privacy by keeping raw data on local devices. The foundational Federated Averaging (FedAvg) algorithm~\cite{mcmahan2017communication} introduced a mechanism where clients train models locally and send only the updates to a central server for aggregation. While effective, FedAvg suffers from high communication costs due to frequent model update exchanges, especially in environments with constrained bandwidth.

To address this challenge, several communication-efficient approaches have been proposed. Lin et al.\cite{lin2017deep} developed Deep Gradient Compression (DGC), which reduces communication overhead through sparsification and momentum correction. Similarly, TernGrad\cite{wen2017terngrad} introduced gradient quantization by transmitting only ternary values, significantly lowering the update size while maintaining model accuracy.

In parallel, efforts have been made to make FL suitable for edge and IoT environments. Wang et al.\cite{wang2019edge} applied FL to Mobile Edge Computing (MEC) to support resource-constrained edge nodes, while Chilukuri and Pesch\cite{chilukuri2020achieving} used FL to predict cache utility dynamically under varying network conditions.

More recently, caching has emerged as a promising technique to further reduce communication overhead in FL systems. Karras et al.\cite{karras2023federated} proposed edge caching of intermediate updates to minimize redundant transmissions. Liu et al.\cite{liu2023cache} allowed clients to update their models in parallel using both cached and newly received global models, reducing latency. Wu et al.~\cite{wu2024fedcache} proposed server-side caching of logits, transmitting only essential outputs to the clients, although this increases client computation.

However, most of these caching strategies are either static or applied only at the client side. In contrast, our proposed FICache introduces an adaptive server-side caching mechanism that selectively defers non-critical updates based on system state. This enables scalable FL in dynamic IoT environments.

\begin{table}[ht]
\centering
\caption{Notation Summary}
\label{tab:notation}
\begin{tabular}{ll}
\hline
\textbf{Symbol} & \textbf{Description} \\
\hline
\( N \) & Total number of clients \\
\( \theta \in \mathbb{R}^d \) & Global model parameters \\
\( t \) & Training round index \\
\( \mathcal{C}^{(t)} \) & Set of participating clients in round \( t \) \\
\( \Delta_i^{(t)} \) & Update from client \( i \) at round \( t \) \\
\( \delta_i^{(t)} \) & Magnitude or utility of \( \Delta_i^{(t)} \) \\
\( \tau \) & Significance threshold for filtering updates \\
\( \text{Cache}_t \) & Set of cached updates at round \( t \) \\
\( C \) & Cache capacity (number of updates) \\
\( \text{Size}(\cdot) \) & Memory size of an update \\
\( \alpha, \beta \) & Weighting factors for accuracy and recency \\
\( \text{Priority}_i \) & Priority score of client \( i \)'s update \\
\( \gamma \) & Priority threshold for update selection \\
\( \mathcal{S}^{(t)} \) & Selected updates for aggregation in round \( t \) \\
\( \mathcal{C}_\text{hit}^{(t)} \) & Clients whose cached updates were reused in round \( t \) \\
\( f(x) \) & ML function predicting best caching strategy \\
\hline
\end{tabular}
\end{table}

\section{System Model and Assumptions}

We consider a standard FL architecture comprising a central server and a set of \( N \) edge clients, each with a private dataset. The server facilitates the aggregation and update process, while clients perform local model updates without sharing raw data.

Let \( \theta \in \mathbb{R}^d \) denote the global model. In each training round \( t \), the server broadcasts \( \theta^{(t)} \) to a subset of participating clients. Each selected client \( i \) computes an update \( \Delta_i^{(t)} \) using its local dataset and transmits it back to the server.

To reduce communication overhead, we introduce a server-side caching mechanism. The server maintains a fixed-size cache of capacity \( C \), which stores a subset of the most valuable client updates. Only updates that pass a significance threshold are considered for aggregation; others may be stored or deferred based on the cache replacement policy.

We assume the following:
\begin{itemize}
    \item Each client has sufficient compute capability to perform local training;
    \item The communication is synchronous across rounds;
    \item Updates vary in quality, and not all are equally beneficial;
    \item Server memory is limited, necessitating replacement strategies such as FIFO, LRU, or PBR.
\end{itemize}

Let \( \mathcal{C}^{(t)} \subseteq \{1, \ldots, N\} \) be the set of clients selected in round \( t \), and let \( \text{Cache}_t \) represent the state of the server-side cache at that time.

\section{System Overview}
\begin{figure}[ht]
    \centering
    \includegraphics[width=0.45\textwidth]{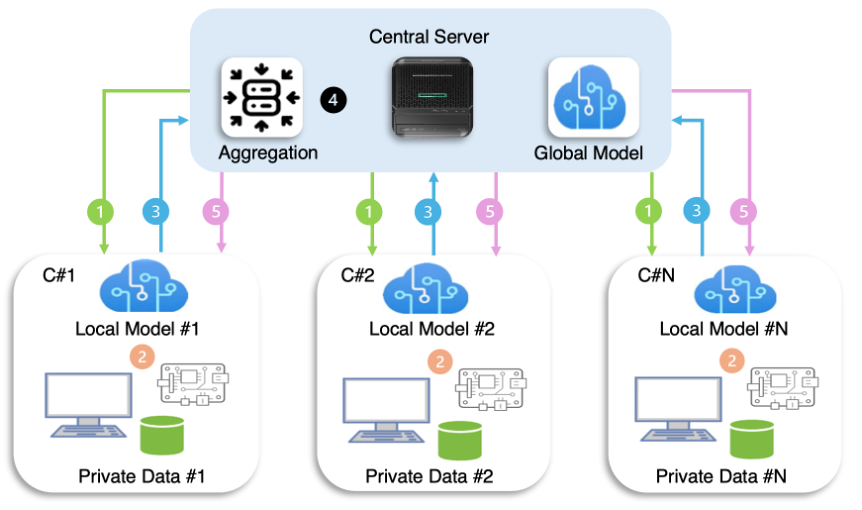}
    \caption{Overview of FL steps inspired by~\cite{mamba2023image}}
    \label{fig:overview}
\end{figure}

This work adopts a FL architecture, as shown in Fig.~\ref{fig:overview}, composed of multiple IoT clients and a central server. Each client independently trains a local model on its private dataset and periodically sends model updates to the server, which performs aggregation to form a global model. This decentralized approach preserves data privacy while enabling collaborative learning across distributed edge devices. The system is implemented using the Flower FL framework~\cite{beutel2020flower}. Experiments are deployed on NVIDIA Jetson Nano~\cite{nvidiaJetsonNano} and Raspberry Pi~\cite{raspberryPi4} devices to simulate constrained edge environments, and on Chameleon Cloud~\cite{chameleon2023} infrastructure for scalability testing.

\vspace{-0.3cm}  

\section{Caching Strategies}

In our proposed framework, the central server maintains a cache to store selected client updates that are deemed significant for aggregation. The goal is to reduce communication overhead by avoiding redundant transmissions of low-impact updates while preserving model accuracy. To manage this cache effectively under memory constraints, we evaluate three classical cache replacement strategies: FIFO, LRU, and PBR.

\begin{figure}[ht]
    \centering
    \includegraphics[width=0.48\textwidth]{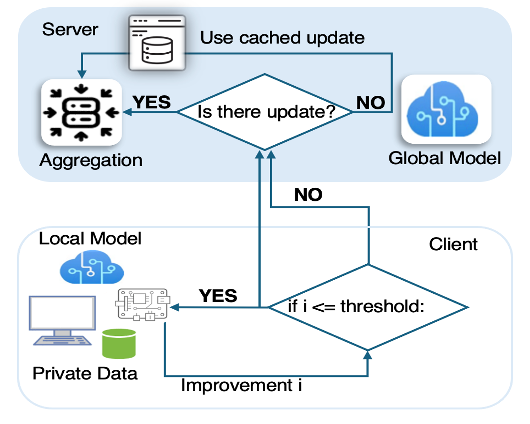}
    \captionsetup{justification=centerlast, singlelinecheck=true}
    \caption{Client-Server Strategy}
    \label{fig:client_server_strategy}
\end{figure}

\subsection{Caching Workflow}

As shown in Fig.~\ref{fig:client_server_strategy}, during each training round \( t \), a client \( i \in \mathcal{C}^{(t)} \) sends its model update \( \Delta_i^{(t)} \) to the server if the local improvement metric exceeds a threshold \( \tau \). Otherwise, the update is withheld or replaced by a cached version from a previous round if available. Formally, an update is transmitted only if:
\[
\delta_i^{(t)} = \|\Delta_i^{(t)}\| \geq \tau,
\]
where \( \delta_i^{(t)} \) represents the magnitude or utility of the client's update. Updates that are not transmitted may still be selected from the cache if they meet inclusion criteria.

\subsection{FIFO Strategy}

The FIFO strategy evicts the oldest cached update when the cache reaches its capacity~\cite{pratheeksha2021mlcache}. It maintains a simple queue of incoming updates:
\begin{itemize}
    \item When \( |\text{Cache}_t| \geq C \), remove the earliest stored update;
    \item Insert the new update at the end of the queue.
\end{itemize}
While computationally efficient, FIFO does not consider update utility or model relevance.

\subsection{LRU Strategy}

The LRU strategy evicts the update that has remained unused for the longest duration~\cite{pratheeksha2021mlcache}:
\begin{itemize}
    \item Each time an update is used in aggregation, it is marked as "recently used";
    \item When cache capacity is full, remove the least recently used update.
\end{itemize}
This approach adapts to client activity patterns and retains frequently used updates.

\subsection{PBR Strategy}

PBR ranks cached updates based on their estimated contribution to model accuracy. For each update \( \Delta_i \), a priority score is computed using:
\[
\text{Priority}_i = \alpha \cdot \text{Accuracy}_i + \beta \cdot \text{Recency}_i,
\]
where \( \alpha \) and \( \beta \) are weighting factors. Updates with the lowest priority are evicted first, allowing the cache to retain high-utility updates.

To guide aggregation and update selection, we define a filtered set of candidate updates as:
\[
\mathcal{S}^{(t)} = \left\{ i \mid \text{Priority}_i \geq \gamma \right\},
\]
where \( \gamma \) is a tunable threshold that determines which updates are retained for aggregation or caching. This selection mechanism ensures that only updates with meaningful contributions are considered under tight memory or communication budgets.

This relevance-aware policy leads to better accuracy–efficiency trade-offs under constrained memory.

\section{Experimental Setup and Evaluation}

To evaluate the effectiveness of our proposed caching framework, we conduct experiments in a realistic FL environment, simulating both resource-constrained edge devices and scalable cloud infrastructure.

\subsection{Testbed and Implementation}


Our FL system is implemented using the Flower framework, which provides flexible support for customizable client-server communication and model training workflows. The experimental setup is deployed across three distinct platforms. Jetson Nano and Raspberry Pi 4 devices are employed to emulate resource-constrained edge clients, reflecting realistic IoT environments. To simulate large-scale client deployments, we utilize the Chameleon Cloud infrastructure, which offers scalable and reconfigurable cloud resources.

To evaluate system behavior, we monitor server memory usage through the \texttt{psutil}\cite{psutil} Python library. Additionally, we capture network traffic using both \texttt{Wireshark}\cite{wireshark} and \texttt{tcpdump}~\cite{tcpdump}, enabling detailed analysis of communication overhead during FL rounds.

\subsection{Datasets}

We evaluate our approach using three datasets. The CIFAR-10 dataset serves as a standard image classification benchmark consisting of 10 distinct classes, commonly used to assess model performance in vision tasks. Additionally, we use medical imaging datasets, namely the Lung and Colon Histopathology Images~\cite{borkowski2019lung}, which are employed for histopathological cancer classification. These datasets help demonstrate the generalizability of our approach across both conventional and domain-specific scenarios.

\subsection{Models}

We evaluate three deep learning architectures to capture a range of computational characteristics across different resource environments. MobileNetV2\cite{howard2017mobilenets} is a lightweight model specifically designed for resource-constrained embedded devices, making it ideal for edge deployment. EfficientNetB0\cite{tan2019efficientnet} represents a balanced architecture that offers an optimal trade-off between accuracy and computational efficiency. In contrast, DenseNet121~\cite{huang2017densely} is a deeper model that achieves high classification accuracy but demands significantly more memory, making it suitable for scenarios where resource availability is less constrained.

\subsection{Caching Configuration}

The server cache size \( C \) is varied across 3, 4, 6, and 8 clients. For each setting, we evaluate three cache replacement strategies: FIFO, LRU, and PBR. A dynamic threshold \( \tau \) is applied to control whether a client update is transmitted. We test thresholds of 1\%, 10\%, and 30\% relative to the improvement magnitude.

\subsection{Evaluation Metrics}

We measure the following:
\begin{itemize}
    \item \textbf{Communication Cost}: total volume of data transmitted during training;
    \item \textbf{Model Accuracy}: average classification accuracy across clients;
    \item \textbf{Memory Usage}: peak server memory used during aggregation.
    \item \textbf{Cache Hits}: the total number of missed client updates that were successfully retrieved from the server-side cache and used during aggregation:
\[
\text{CacheHits}_T = \sum_{t=1}^T |\mathcal{C}_\text{hit}^{(t)}|,
\]
where \( \mathcal{C}_\text{hit}^{(t)} \) is the set of clients in round \( t \) whose updates were not transmitted but were retrieved from the cache.

\end{itemize}

\section{Results and Analysis}

This section presents a detailed analysis of our experimental results across three key dimensions: communication cost, model accuracy, and memory efficiency. We compare the baseline FedAvg configuration with our caching-based approach under various thresholds and replacement strategies.

\subsection{Communication Cost Reduction}

Caching significantly reduces communication overhead by filtering and reusing updates. We define the total communication cost over \( T \) rounds as:
\[
\text{CommCost}_T = \sum_{t=1}^{T} \sum_{i \in \mathcal{C}^{(t)}} \mathbf{1}[\delta_i^{(t)} \geq \tau] \cdot \text{Size}(\Delta_i^{(t)}),
\]
where \( \mathcal{C}^{(t)} \) is the set of participating clients in round \( t \), \( \delta_i^{(t)} \) is the magnitude of the client update, and \( \tau \) is the threshold. The indicator function \( \mathbf{1}[\delta_i^{(t)} \geq \tau] \) ensures that only significant updates (above threshold) are counted, and \( \text{Size}(\Delta_i^{(t)}) \) is the size of the transmitted update.

\begin{figure}[ht]
    \centering
    \includegraphics[width=0.48\textwidth]{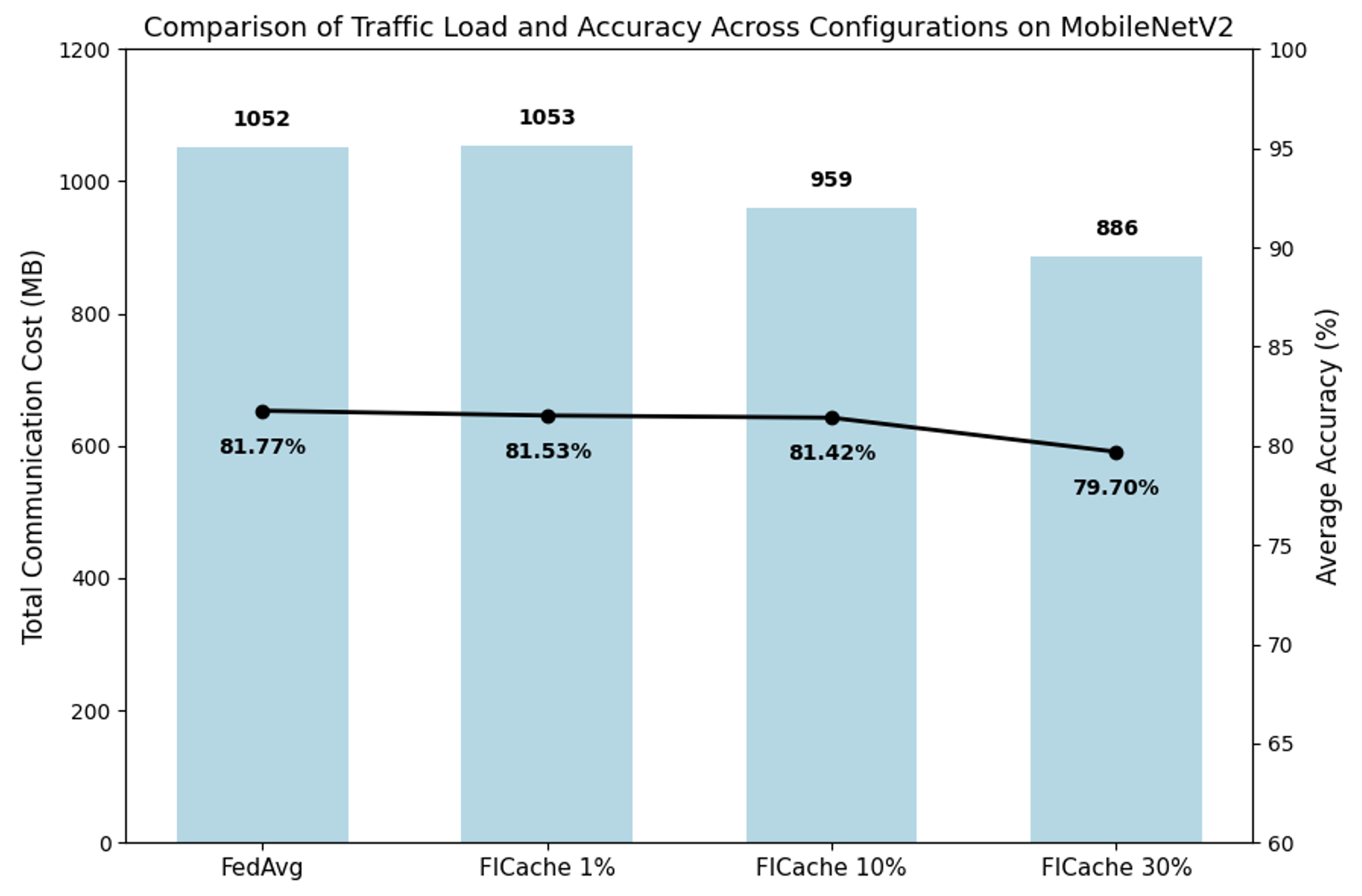}
    \caption{Comparison of total communication cost and accuracy across different threshold configurations using MobileNetV2.}
    \label{fig:threshold_tradeoff}
\end{figure}

As shown in Figure~\ref{fig:threshold_tradeoff}, using a 30\% threshold with MobileNetV2 on CIFAR-10 lowers the total communication volume from 1052 MB to 886 MB—a reduction of over 15\%.

Lower thresholds (e.g., 1\%) allow more updates to be sent, improving learning speed but increasing communication. Our results confirm that threshold-based caching offers a tunable balance between efficiency and update frequency.

\subsection{Model Accuracy}

\begin{figure}[ht]
    \centering
    \includegraphics[width=0.48\textwidth]{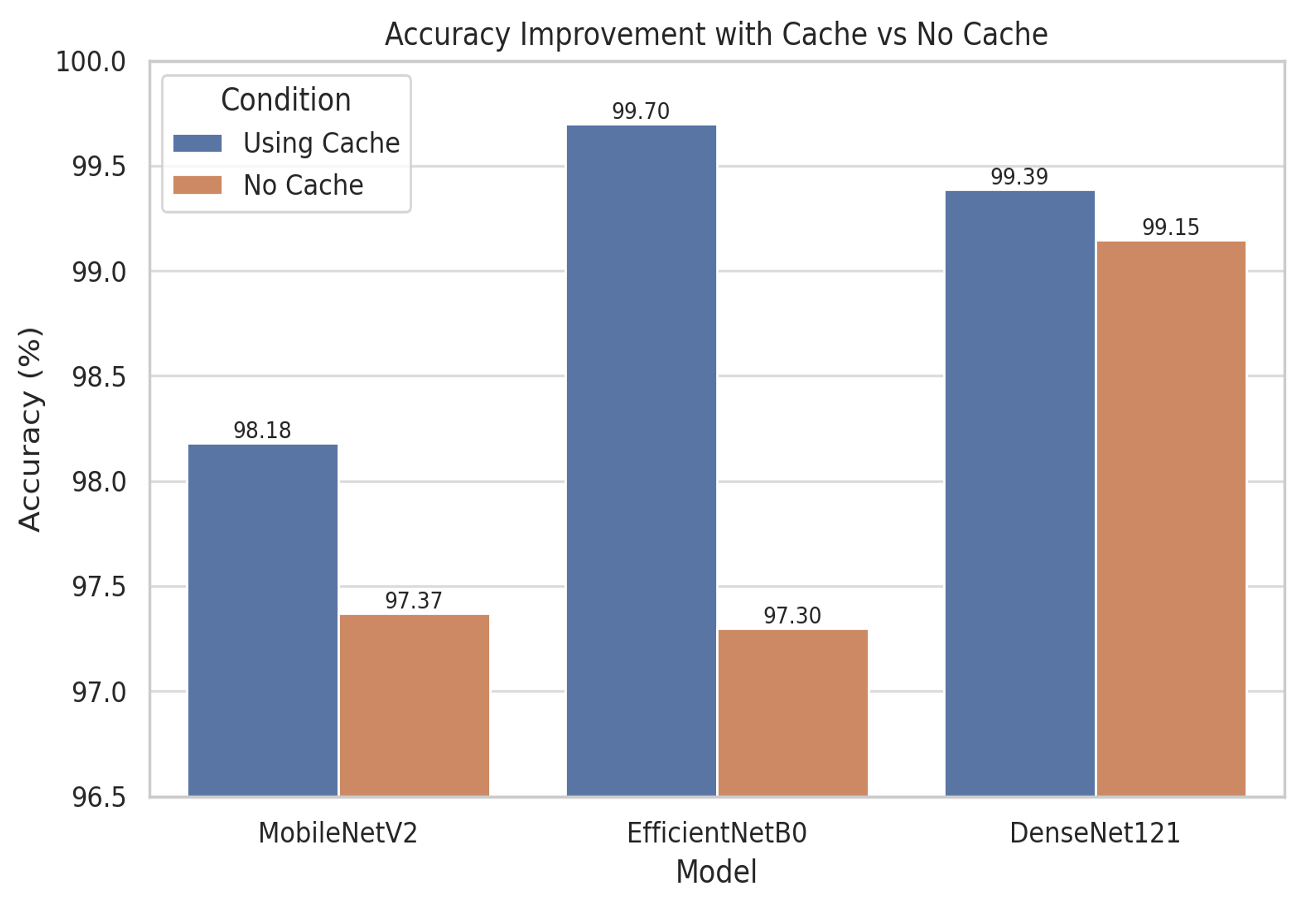}
    \caption{Accuracy improvement with cache vs. no cache across three models.}
    \label{fig:accuracy_comparison}
\end{figure}

Despite the reduction in communication, model accuracy is preserved—or even improved—when caching is enabled. As illustrated in Figure~\ref{fig:accuracy_comparison}, MobileNetV2 shows an increase in accuracy from 97.37\% to 98.18\%, while EfficientNetB0 achieves a more substantial improvement of 2.4\%, rising from 97.30\% to 99.70\%. Similarly, DenseNet121 gains 0.24\%, improving from 99.15\% to 99.39\%. These enhancements are largely attributed to the reuse of previously effective client updates that would otherwise have been discarded for falling below the update threshold.

\subsection{Memory Efficiency}

We define the total memory usage of the server cache at round \( t \) as:
\[
\text{MemUsage}_t = \sum_{j=1}^{|\text{Cache}_t|} \text{Size}(\Delta_j),
\]
where \( \text{Cache}_t \) is the set of cached updates at round \( t \), and \( \text{Size}(\Delta_j) \) denotes the memory size of update \( j \).

\begin{figure}[ht]
    \centering
    \includegraphics[width=0.48\textwidth]{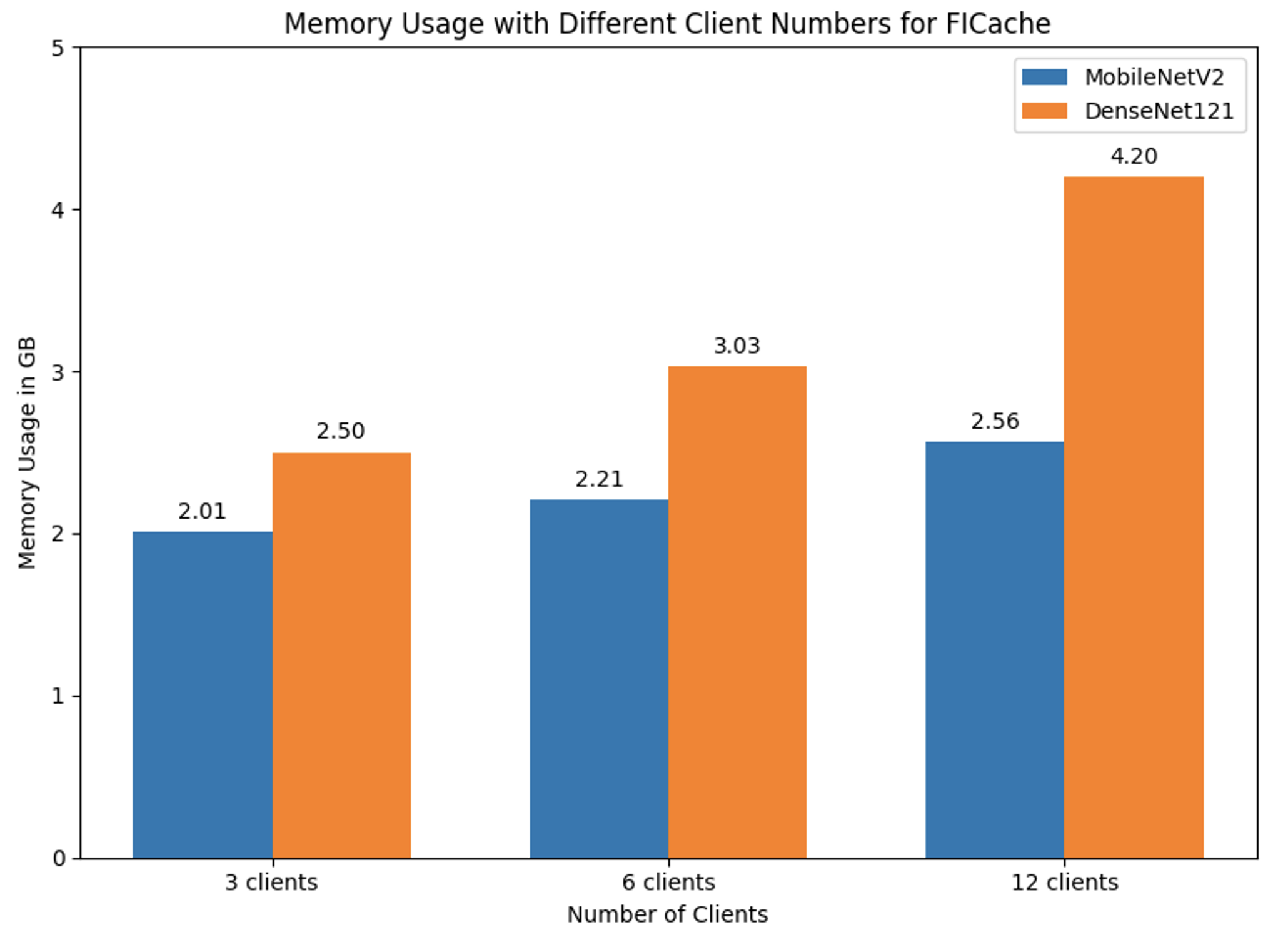}
    \caption{Memory usage for FICache across 3, 6, and 12 clients.}
    \label{fig:mem_usage_12clients}
\end{figure}

This formulation helps quantify the impact of caching on edge devices with limited memory. Figure~\ref{fig:mem_usage_12clients} compares the memory usage of MobileNetV2 and DenseNet121 when caching is applied across different numbers of clients (3, 6, and 12). DenseNet121 consistently consumes more memory than MobileNetV2 across all settings.

For example, with 3 clients, MobileNetV2 uses 2.01 GB, while DenseNet121 reaches 2.50 GB. At 6 clients, memory usage increases to 2.21 GB for MobileNetV2 and 3.03 GB for DenseNet121. With 12 clients, DenseNet121 exceeds 4.20 GB, surpassing the 3.87 GB memory limit of Jetson Nano devices. In contrast, MobileNetV2 remains under the limit at 2.56 GB.

These results highlight the need for intelligent caching and client selection to prevent memory overload and ensure stable operation under varying client loads. As expected, memory consumption increases with the number of clients, further emphasizing the importance of scalable memory management.

\subsection{Impact of Cache Replacement Strategies}

To evaluate the effectiveness of different cache replacement strategies in FL, we used supervised learning models to predict the most suitable strategy—FIFO, LRU, or PBR—based on key system features such as model type, dataset size, cache capacity, threshold, and data distribution. This data-driven approach provides valuable insights into which strategy is likely to perform best under various deployment scenarios.

\begin{figure}[ht]
    \centering
    \includegraphics[width=0.45\textwidth]{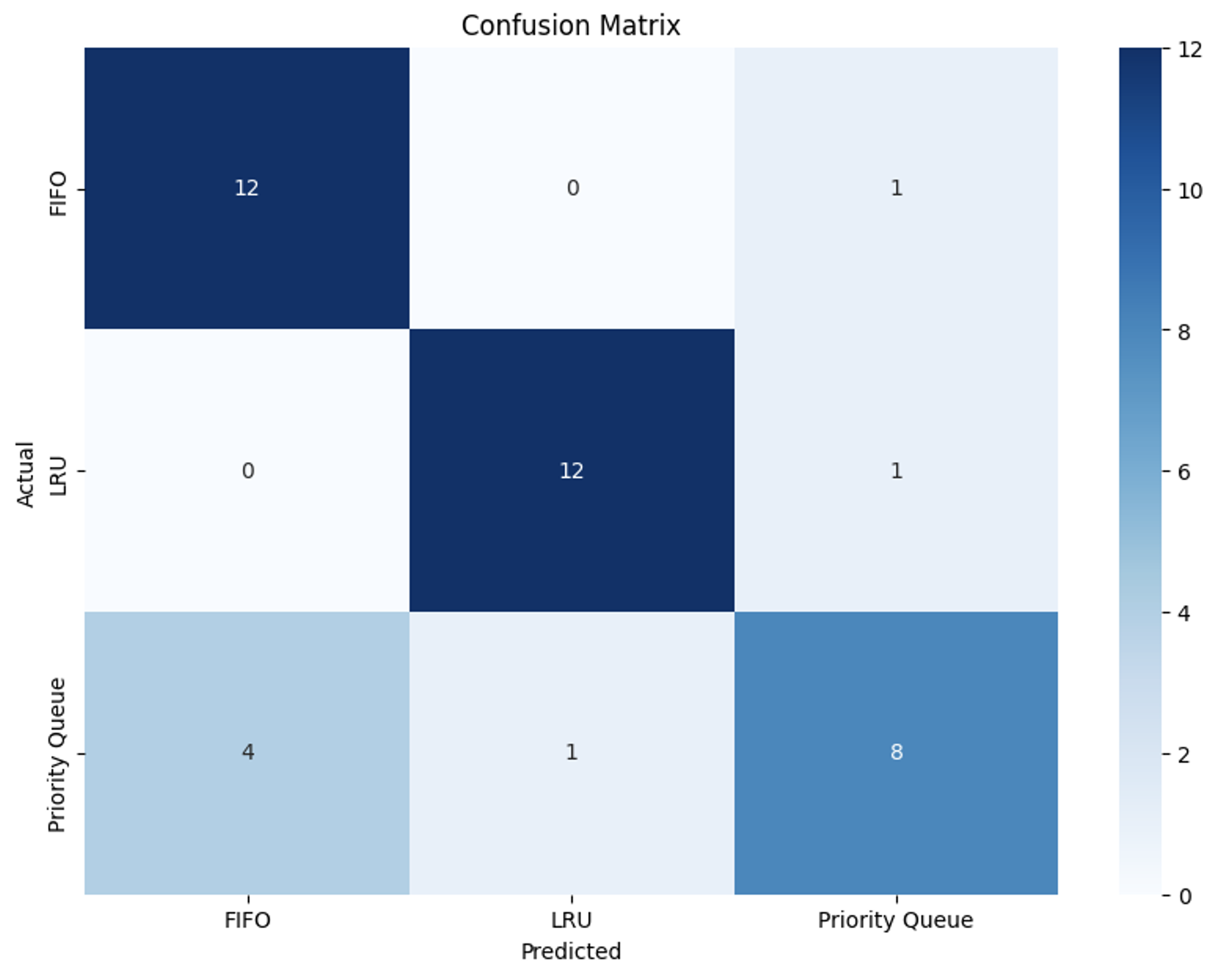}
    \caption{Confusion matrix of classifier predictions across caching strategies.}
    \label{fig:confusion_matrix}
\end{figure}

Figure~\ref{fig:confusion_matrix} presents the confusion matrix of classifier predictions using XGBoost across caching strategies.


Overall, our caching framework offers a robust way to reduce communication, maintain accuracy, and support edge deployment under resource constraints.

\section{Conclusion}

In this work, we proposed a caching-based communication optimization framework for FL in resource-constrained IoT environments. By filtering insignificant updates and reusing valuable ones through server-side caching, our approach significantly reduces communication cost while preserving or enhancing model accuracy.

We implemented and evaluated three cache replacement strategies—FIFO, LRU, and PBR across multiple datasets and deep learning models. Experimental results demonstrate that our method reduces total communication by up to 20\%, maintains high model accuracy, and improves memory efficiency on edge devices such as Jetson Nano and Raspberry Pi.

This work contributes to scalable FL deployment in real-world settings like smart cities and healthcare, where bandwidth and memory are limited.




\bibliographystyle{IEEEtran}
\bibliography{references}

\begin{thebibliography}{10}
\providecommand{\url}[1]{#1}
\csname url@samestyle\endcsname
\providecommand{\newblock}{\relax}
\providecommand{\bibinfo}[2]{#2}
\providecommand{\BIBentrySTDinterwordspacing}{\spaceskip=0pt\relax}
\providecommand{\BIBentryALTinterwordstretchfactor}{4}
\providecommand{\BIBentryALTinterwordspacing}{\spaceskip=\fontdimen2\font plus
\BIBentryALTinterwordstretchfactor\fontdimen3\font minus \fontdimen4\font\relax}
\providecommand{\BIBforeignlanguage}[2]{{%
\expandafter\ifx\csname l@#1\endcsname\relax
\typeout{** WARNING: IEEEtran.bst: No hyphenation pattern has been}%
\typeout{** loaded for the language `#1'. Using the pattern for}%
\typeout{** the default language instead.}%
\else
\language=\csname l@#1\endcsname
\fi
#2}}
\providecommand{\BIBdecl}{\relax}
\BIBdecl

\bibitem{mcmahan2017communication}
B.~McMahan, E.~Moore, D.~Ramage, S.~Hampson, and B.~A. y~Arcas, ``Communication-efficient learning of deep networks from decentralized data,'' in \emph{Artificial intelligence and statistics}.\hskip 1em plus 0.5em minus 0.4em\relax PMLR, 2017, pp. 1273--1282.

\bibitem{pratheeksha2021mlcache}
\BIBentryALTinterwordspacing
P.~Pratheeksha and S.~A. Revathi, ``Machine learning-based cache replacement policies: A survey,'' \emph{International Journal of Engineering and Advanced Technology (IJEAT)}, vol.~10, no.~6, pp. 19--22, August 2021. [Online]. Available: \url{https://doi.org/10.35940/ijeat.F2907.0810621}
\BIBentrySTDinterwordspacing

\bibitem{krizhevsky2009learning}
A.~Krizhevsky, G.~Hinton \emph{et~al.}, ``Learning multiple layers of features from tiny images,'' 2009.

\bibitem{lin2017deep}
Y.~Lin, S.~Han, H.~Mao, Y.~Wang, and W.~J. Dally, ``Deep gradient compression: Reducing the communication bandwidth for distributed training,'' \emph{arXiv preprint arXiv:1712.01887}, 2017.

\bibitem{wen2017terngrad}
W.~Wen, C.~Xu, F.~Yan, C.~Wu, Y.~Wang, Y.~Chen, and H.~Li, ``Terngrad: Ternary gradients to reduce communication in distributed deep learning,'' \emph{Advances in neural information processing systems}, vol.~30, 2017.

\bibitem{wang2019edge}
X.~Wang, Y.~Han, C.~Wang, Q.~Zhao, X.~Chen, and M.~Chen, ``In-edge ai: Intelligentizing mobile edge computing, caching and communication by federated learning,'' \emph{Ieee Network}, vol.~33, no.~5, pp. 156--165, 2019.

\bibitem{chilukuri2020achieving}
S.~Chilukuri and D.~Pesch, ``Achieving optimal cache utility in constrained wireless networks through federated learning,'' in \emph{2020 IEEE 21st International Symposium on" A World of Wireless, Mobile and Multimedia Networks"(WoWMoM)}.\hskip 1em plus 0.5em minus 0.4em\relax IEEE, 2020, pp. 254--263.

\bibitem{karras2023federated}
A.~Karras, C.~Karras, K.~Giotopoulos, D.~Tsolis, K.~Oikonomou, and S.~Sioutas, ``Federated edge intelligence and edge caching mechanisms. information 2023, 14, 414,'' 2023.

\bibitem{liu2023cache}
Y.~Liu, L.~Su, C.~Joe-Wong, S.~Ioannidis, E.~Yeh, and M.~Siew, ``Cache-enabled federated learning systems,'' in \emph{Proceedings of the Twenty-fourth International Symposium on Theory, Algorithmic Foundations, and Protocol Design for Mobile Networks and Mobile Computing}, 2023, pp. 1--11.

\bibitem{wu2024fedcache}
Z.~Wu, S.~Sun, Y.~Wang, M.~Liu, K.~Xu, W.~Wang, X.~Jiang, B.~Gao, and J.~Lu, ``Fedcache: A knowledge cache-driven federated learning architecture for personalized edge intelligence,'' \emph{IEEE Transactions on Mobile Computing}, vol.~23, no.~10, pp. 9368--9382, 2024.

\bibitem{beutel2020flower}
D.~J. Beutel, T.~Topal, A.~Mathur, X.~Qiu, J.~Fernandez-Marques, Y.~Gao, L.~Sani, K.~H. Li, T.~Parcollet, P.~P.~B. de~Gusm{\~a}o \emph{et~al.}, ``Flower: A friendly federated learning research framework,'' \emph{arXiv preprint arXiv:2007.14390}, 2020.

\bibitem{nvidiaJetsonNano}
{NVIDIA}, ``Jetson nano developer kit,'' \url{https://developer.nvidia.com/embedded/jetson-nano}, n.d., accessed: 2025-06-28.

\bibitem{raspberryPi4}
{Raspberry Pi Foundation}, ``Raspberry pi 4 model b,'' \url{https://www.raspberrypi.com/products/raspberry-pi-4-model-b/}, n.d., accessed: 2025-06-28.

\bibitem{chameleon2023}
{Chameleon Cloud}, ``Chameleon: A large-scale, reconfigurable experimental environment for cloud research,'' \url{https://chameleoncloud.io/}, 2023, accessed: 2025-06-28.

\bibitem{mamba2023image}
D.~Mamba~Kabala, A.~Hafiane, L.~Bobelin, and R.~Canals, ``Image-based crop disease detection with federated learning,'' \emph{Scientific Reports}, vol.~13, no.~1, p. 19220, 2023.

\bibitem{psutil}
{Psutil}, ``Python system and process utilities,'' \url{https://psutil.readthedocs.io/}, n.d., accessed: 2025-06-29.

\bibitem{wireshark}
{Wireshark}, ``Wireshark: Go deep,'' \url{https://www.wireshark.org/}, n.d., accessed: 2025-06-29.

\bibitem{tcpdump}
{Tcpdump \& libpcap}, ``Tcpdump \& libpcap,'' \url{http://www.tcpdump.org/}, n.d., accessed: 2025-06-29.

\bibitem{borkowski2019lung}
A.~A. Borkowski, M.~M. Bui, L.~B. Thomas, C.~P. Wilson, L.~A. DeLand, and S.~M. Mastorides, ``Lung and colon cancer histopathological image dataset (lc25000),'' \emph{arXiv preprint arXiv:1912.12142}, 2019.

\bibitem{howard2017mobilenets}
A.~G. Howard, M.~Zhu, B.~Chen, D.~Kalenichenko, W.~Wang, T.~Weyand, M.~Andreetto, and H.~Adam, ``Mobilenets: Efficient convolutional neural networks for mobile vision applications,'' \emph{arXiv preprint arXiv:1704.04861}, 2017.

\bibitem{tan2019efficientnet}
M.~Tan and Q.~Le, ``Efficientnet: Rethinking model scaling for convolutional neural networks,'' in \emph{International conference on machine learning}.\hskip 1em plus 0.5em minus 0.4em\relax PMLR, 2019, pp. 6105--6114.

\bibitem{huang2017densely}
G.~Huang, Z.~Liu, L.~Van Der~Maaten, and K.~Q. Weinberger, ``Densely connected convolutional networks,'' in \emph{Proceedings of the IEEE conference on computer vision and pattern recognition}, 2017, pp. 4700--4708.

\end{thebibliography}

\vfill

\end{document}